# Data Analysis of Bright Main-Sequence A- and B-type Stars Observed Using the TESS and BRITE Spacecraft

*Joyce A. Guzik[1], Jason Jackiewicz[2], Andrzej Pigulski[3], Giovanni Catanzaro[4],
Michael S. Soukup[5], Patrick Gaulme[6], Gerald Handler[7], and the BRITE Team*

**Abstract**

During the last two years we have received long time-series photometric observations of bright (V mag < 8) main-sequence A- and B-type stars observed by the NASA TESS spacecraft[8] and the Austria-Poland-Canada BRITE[9] satellites.  Using TESS observations of metallic-line A (Am) stars having peculiar element abundances, our goal is to determine whether and why these stars pulsate in multiple radial and nonradial modes, as do the $\delta$ Scuti stars in the same region of the H-R diagram.  The BRITE data were requested to investigate pulsations in bright (V around 6 mag) A- and B-type stars in the Cygnus-Lyra field of view that had been proposed for observations during the now-retired NASA *Kepler* mission[10].

Of the 21 (out of 62 proposed) Am stars observed by TESS so far, we find one $\delta$ Sct star and two $\delta$ Sct / $\gamma$ Dor hybrid candidates.  Of the remaining stars, we find three $\gamma$ Dor candidates, six stars showing photometric variations that may or may not be associated with pulsations, and eight stars without apparent significant photometric variability.  For the A- and B-type stars observed by BRITE, one star (HR 7403) shows low amplitude low frequency modes that likely are associated with its B(emission) star properties; one star (HR 7179) shows SPB variability that is also found in prior *Kepler* data, and two stars (HR 7284 and HR 7591) show no variability in BRITE data, although very low amplitude variability was found in TESS or *Kepler* data. For the TESS and BRITE targets discussed here, follow-up ground- and space-based photometric and spectroscopic observations combined with stellar modeling will be needed to constrain stellar parameters and to understand the nature of the variability.

## 1. Introduction

Stars of nearly every type and evolutionary state show pulsational variability (Aerts, Christensen-Dalsgaard, and Kurtz 2010). The pulsation properties can be used to infer the internal structure and processes in stars, and validate stellar model physics and theoretical interpretations. This paper focuses on several types of main-sequence (core hydrogen-burning) variable stars that pulsate in one or more radial and nonradial pulsation modes, namely the $\gamma$ Doradus, $\delta$ Scuti, Slowly Pulsating B-type (SPB) stars, and $\beta$ Cephei variables.  Greek letters, such as $\beta$ = beta, $\gamma$ = gamma, and $\delta$ = delta, are used to refer to the brightest stars in a constellation in order of the Greek alphabet, and many variable star types are named after their prototype star.

Here we show results of searches for pulsations using photometric data obtained by the Transiting Exoplanet Survey Satellite (TESS) as part of their Cycle 2 Guest Investigator Program, and by the BRITE-Constellation satellites.

---

[1] Los Alamos National Laboratory, MS T082, Los Alamos, NM 87547, joy@lanl.gov
[2] Department of Astronomy, New Mexico State University, Las Cruces, NM
[3] Instytut Astronomiczny, Uniwersytet Wrocławski, Wrocław, Poland
[4] INAF--Osservatorio Astrofisico di Catania, Via S. Sofia 78, I-95123 Catania, Italy
[5] Los Alamos National Laboratory (retired), Albuquerque, NM 87111
[6] Max-Planck-Institut für Sonnensystemforschung, Justus-von-Liebig-Web 3, 37077, Göttingen, Germany
[7] Nicolaus Copernicus Astronomical Center, Bartycka 18, 00-716 Warsaw, Poland

[8] See https://www.nasa.gov/tess-transiting-exoplanet-survey-satellite and https://tess.mit.edu/

[9] See https://brite-constellation.at   Based on data collected by the BRITE-Constellation satellite mission, designed, built, launched, operated and supported by the Austrian Research Promotion Agency (FFG), the University of Vienna, the Technical University of Graz, the University of Innsbruck, the Canadian Space Agency (CSA), the University of Toronto Institute for Aerospace Studies (UTIAS), the Foundation for Polish Science & Technology (FNiTP MNiSW), and National Science Centre (NCN).

[10] See https://www.nasa.gov/mission_pages/kepler/main/index.html

## 2. Am Star Results from TESS Guest Investigator Program

The NASA TESS spacecraft was launched on April 18, 2018 and is in an elliptical 13.7-day lunar resonance orbit around Earth. Its main mission is to search for transiting exoplanets around nearby G, K, and M spectral type stars using the transit method. The TESS spacecraft has four CCD cameras aligned in a vertical strip to view a 24° by 90° section of the sky (called a 'sector') continuously for 27 days before moving to the next (partially overlapping) sector. The sky below the ecliptic plane (negative declination) was observed in 13 sectors for the first year of operation (Cycle 1), and observations are in progress for 13 sectors at positive declination above the ecliptic plane (Cycle 2). Because the sectors overlap, targets nearer the north and south ecliptic poles will be observable for up to a year. Data products include full-frame images with a 30-minute cadence, as well as 2-min cadence light curves for several hundred thousand stars in each sector. As of this writing (May 2020) TESS data is available from the Mikulski Archive for Space Telescopes (MAST)[11] for sectors 1 through 23; this paper includes analysis of observations through sector 22.

Catanzaro et al. (2019) used high-resolution spectroscopy + Gaia DR2 parallaxes to determine stellar parameters and detailed abundances for 62 metallic-line A (Am) stars. These stars show significant underabundances of Ca and Sc, and enhanced abundances of Ti and Fe-group elements. Catanzaro et al. showed that many of these stars are located in the δ Sct and δ Sct/γ Dor hybrid regions of the Hertzsprung-Russell (H-R) diagram. For Cycle 2 of the TESS Guest Investigator Program, we requested two-minute cadence observations of these stars[12], most of which have magnitudes 7-8, to determine whether these stars pulsate.

The δ Sct variables are main-sequence variables about twice the mass of the Sun and pulsate in multiple radial and nonradial modes with periods of about 2 hours, or frequencies about 12 cycles/day (Aerts et al. 2010, Breger 2000). The cause of their pulsations is believed to be the 'kappa', or opacity valving mechanism, named for the Greek letter κ (kappa) used to represent opacity. Opacity is a measure of how 'opaque' or resistant the stellar interior plasma is to transfer of photon radiation through the layer. Figure 1 shows the opacity versus interior temperature from a 2 solar mass model. For the δ Sct stars, the increased opacity region labeled He+ where the second electron of helium is ionizing at 50,000 K in the stellar envelope is at an optimum location to block radiation, causing this layer to absorb heat, expand, and then cool and contract in a feedback loop to produce the pulsations. For the more massive β Cep and SPB stars that we discuss later, the opacity "Z-bump" at 200,000 K, where electron transitions in iron-group elements are inhibiting radiation transfer, is at the right location in the stellar envelope to cause κ-effect pulsation driving.

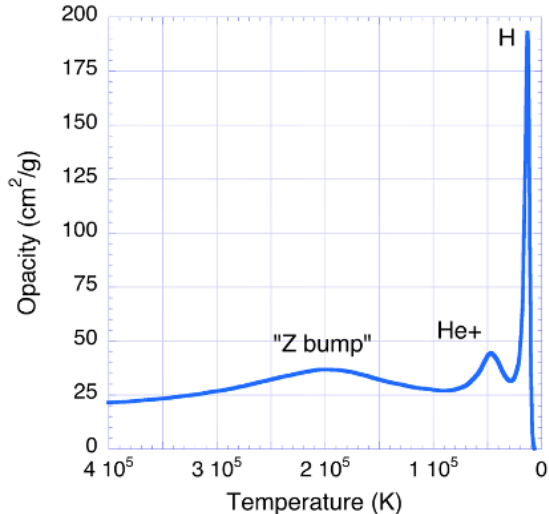

**Figure 1: Opacity vs. interior temperature for 2 solar-mass main-sequence stellar model. In this plot, the stellar interior corresponds to higher temperatures to left of the x-axis, while the stellar surface is toward the right. The opacity enhancement at around 50,000 K (labeled He+) where the second electron of helium is being ionized is responsible for driving δ Sct pulsations. In the more massive SPB and β Cep stars, the opacity increase at around 200,000 K (labeled 'Z bump') caused by electronic transitions in Fe-group elements is responsible for driving pulsations.**

The γ Dor variables are about 1.5 times the mass of the Sun and pulsate in nonradial gravity modes with periods of 1 to 3 days (Kaye et al. 1999). The convective blocking mechanism at the base of the envelope convection zone (Guzik et al. 2000) is proposed to drive the pulsations. The pulsation driving arises because the pulsation period is shorter than the local convective timescale at the base of the envelope convection zone, and so the convection cannot adapt quickly enough during a pulsation cycle to transport the radiation emerging from the stellar center, causing radiation flow to be blocked periodically.

---

[11] https://archive.stsci.edu/

[12] See Guzik et al. TESS GI program G022027 at https://heasarc.gsfc.nasa.gov/docs/tess/approved-programs.html

Before the high-precision long time-series photometric observations made by the *Kepler* spacecraft (Borucki et al. 2010; Gilliland et al. 2010), theory and stellar models for the most part explained the locations of the δ Sct and γ Dor instability regions in the H-R diagram, and hybrid variables pulsating in both low- and high-frequency modes were found in the overlapping region between these two instability regions (see Fig. 16). However, the *Kepler* data showed hybrid stars scattered throughout both instability regions, and even beyond the edges (Grigahcene et al. 2010; Uytterhoeven et al. 2011). Therefore, we do not necessarily expect that pulsators found in the TESS Am star data will be located in the pre-*Kepler* defined instability regions.

For the Am stars, the diffusive processes responsible for the abundance peculiarities are also expected to cause helium to settle out of the pulsating driving region, making the δ Sct pulsation driving mechanism ineffective (Breger 1970; Kurtz 1976). Nevertheless, some Am stars do show pulsations, raising questions about whether another pulsation mechanism is responsible (Smalley et al. 2017). Other pulsation driving mechanisms also are being explored to explain the prevalence of hybrid stars throughout the γ Dor and δ Sct instability regions.

## 2.1 TESS Light Curves and Amplitude Spectra

So far, 21 of the 62 Am stars in the Catanzaro et al. (2019) sample have been observed in 2-min cadence during TESS Cycle 2, many during observations in more than one sector. The data is publicly available from the MAST database. Below we show some of the light curves and amplitude spectra (produced using a Fourier transform of the time series data). Figures 2-5 show results for hybrid candidates, Fig. 6 shows the spectrum of a δ Sct star, Figs. 7-10 show results for γ Dor candidates, Fig. 11 shows the light curve of an eclipsing binary, and Figs. 12-15 show light curves with large amplitude variations (4 to 15 parts per thousand) and only one or at most two significant periodicities in the amplitude spectra that may be caused by rotation and magnetic activity.

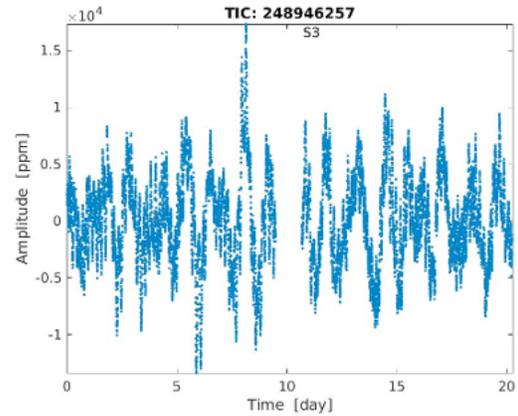

**Figure 2: TESS light curve for TIC 248946257 observed in sector 3.**

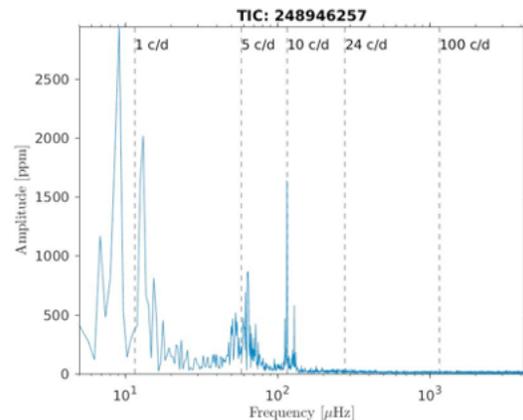

**Figure 3: TESS amplitude spectrum of TIC 248946257, a hybrid δ Sct/γ Dor candidate. Both low frequency γ Dor pulsations (around 1 cycle/day) and higher frequency δ Sct pulsations (near 5 and 10 cycles/day) are apparent.**

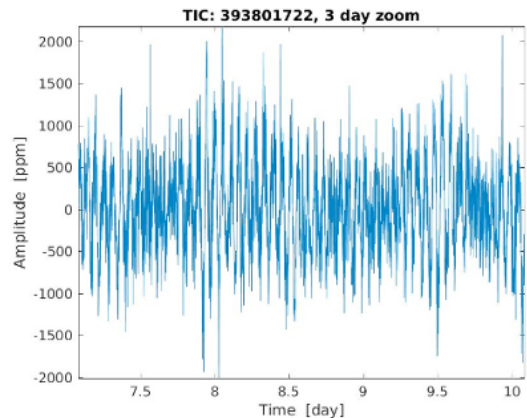

**Figure 4: Three-day zoom-in of TESS light curve for TIC 393801722 observed in sector 22.**

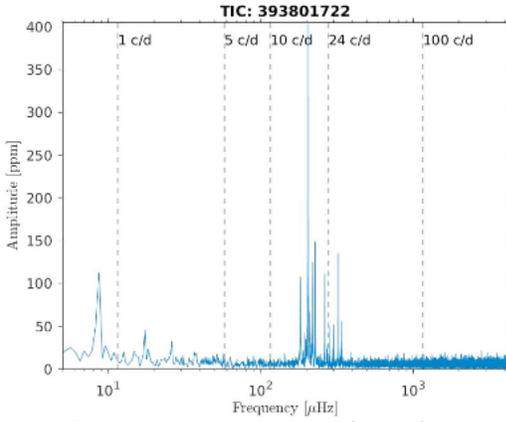

**Figure 5:** Amplitude spectrum for TIC 393801722, revealed as a hybrid δ Sct / γ Dor candidate.

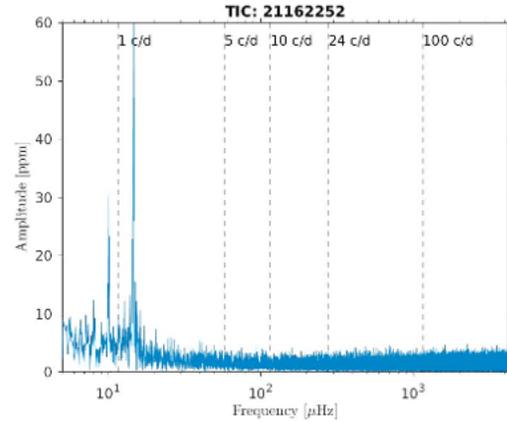

**Figure 8:** Amplitude spectrum of TIC 21162252, revealed as a γ Dor candidate.

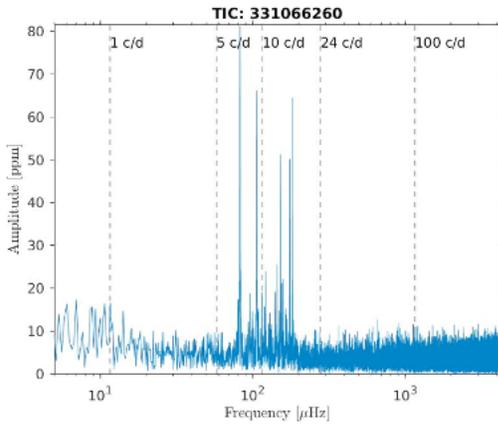

**Figure 6:** Amplitude spectrum of TIC 331066260, a low-amplitude δ Sct star observed by TESS in sectors 16 and 17.

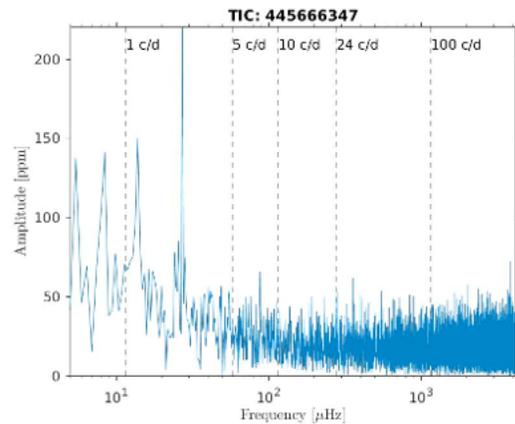

**Figure 9:** Amplitude spectrum of TIC 445666347, revealed as a γ Dor candidate, observed by TESS in sector 18.

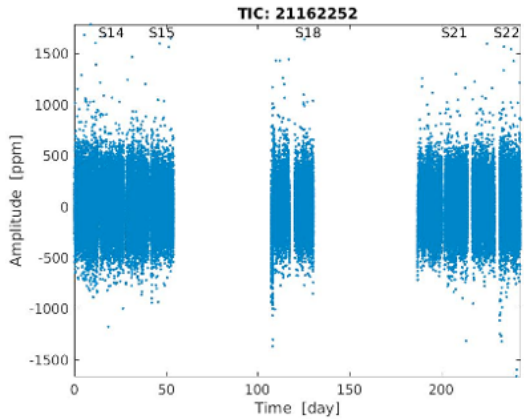

**Figure 7:** TESS light curve for TIC 21162252.

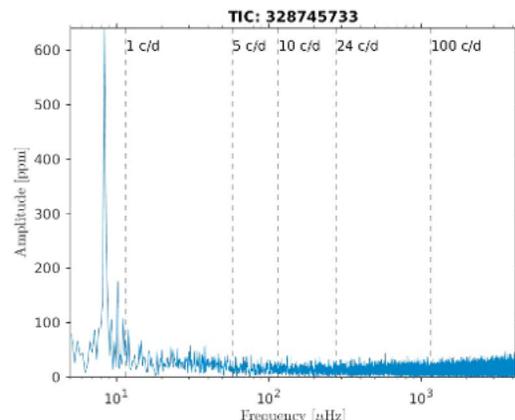

**Figure 10:** Amplitude spectrum of TIC 328745733, revealed as a γ Dor candidate, observed in by TESS in sectors 16 and 17.

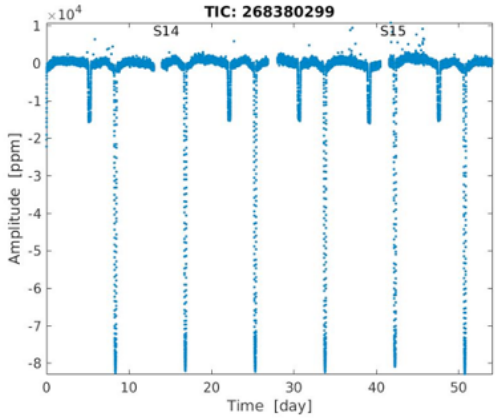

Figure 11: TESS light curve of TIC 268380299, a previously known eclipsing binary with a γ Dor component discovered using *Kepler* data (Çakırlı 2015).

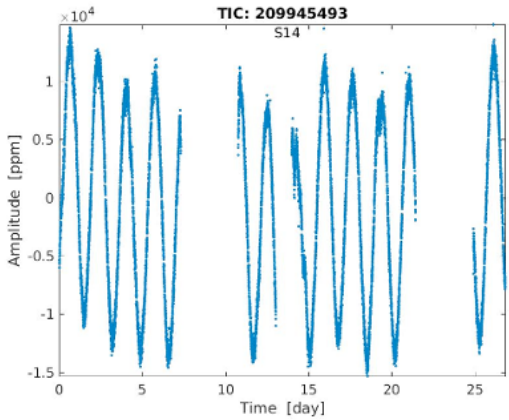

Figure 12: TESS light curve for TIC 209945493, showing large amplitude variation.

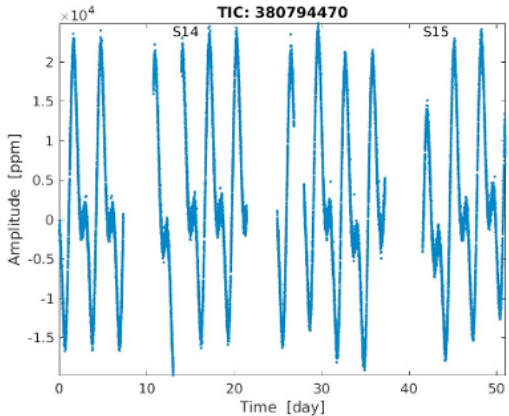

Figure 13: TESS light curve for TIC 380794470, showing large amplitude variation.

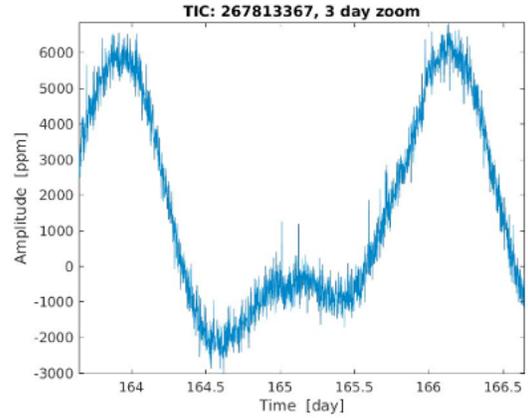

Figure 14: 3-day zoom-in of TESS light curve for TIC 267813367, showing large amplitude variation. Data are available from sectors 15, 16, 17, 20, and 21.

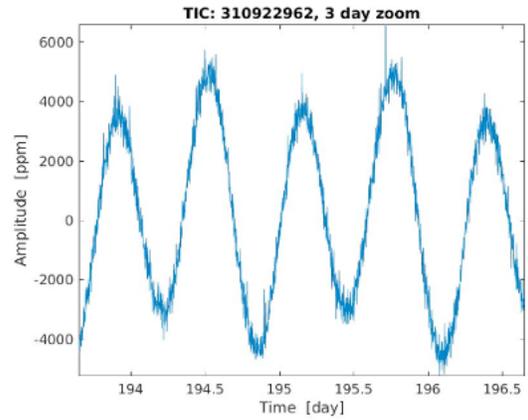

Figure 15: 3-day zoom-in of TESS light curve for TIC 310922962, showing large amplitude variation. Data are available from sectors 14, 15, and 21.

## 2.2 Summary of TESS Results

Table 1 lists the 21 stars from the Catanzaro et al. sample observed by TESS to date, along with their TESS Input Catalog (TIC) numbers, HD catalog number, TESS magnitude, effective temperature and luminosity as derived by Catanzaro et al. (2019), variable-star type prediction based on the star's location in the H-R diagram relative to the instability strip boundaries, and the variable-star type results from TESS data.

Out of the 21 Am stars observed so far by TESS, we find: two hybrid candidates, one δ Sct star, and three γ Dor (candidates). The eclipsing binary, also known as V2094 Cyg, was previously discovered from *Kepler* data to have a γ Dor component (Çakırlı 2015). We also find eight 'constant' stars showing no apparent photometric variability, as well as four stars with high amplitude (4-15 parts per thousand) variations and two with low amplitude variations of unknown origin. These variations may be due to rotation and magnetic activity, or actually may be γ Dor pulsation modes, or may be the result of the orbital motion of a close interacting binary. The binary interpretation is doubtful because Catanzaro et al. (2019) screened the sample to remove spectroscopic binaries. Because the rotation periods and γ Dor pulsation periods can be similar, it is difficult to distinguish between these possibilities, which is the reason that the γ Dor and hybrid stars are referred to as 'candidates'.

Table 1. Summary of TESS GI results through sector 22. The $T_{eff}$, log $L/L_{sun}$, and variable star type prediction based on H-R Diagram location are from Catanzaro et al. (2019).

| HD Number | TIC ID | TESS Mag. | $T_{eff} \pm$ 125 K | log $L/L_{sun}$ | Prediction | Result |
|---|---|---|---|---|---|---|
| HD 10088 | 151056397 | 7.594 | 7700 | 0.81±0.1 | δ Sct | Constant |
| HD 14825 | 445666347 | 7.729 | 8500 | 1.23±0.14 | δ Sct | γ Dor candidate |
| HD 99620 | 310922962 | 7.501 | 7700 | 0.93±0.1 | δ Sct | High ampl. variation |
| HD 108449 | 393801722 | 8.033 | 7000 | 0.89±0.12 | Hybrid | Hybrid candidate |
| HD 127263 | 158035384 | 7.916 | 8200 | 0.88±0.18 | δ Sct | Constant |
| HD 143914 | 313423751 | 7.709 | 8000 | 0.96±0.1 | δ Sct | Constant |
| HD 149650 | 198210258 | 5.843 | 8800 | 1.64±0.14 | None | Low ampl. variation |
| HD 149748 | 198212148 | 7.072 | 7600 | 0.77±0.1 | δ Sct | Constant |
| HD 164394 | 329343467 | 7.281 | 7700 | 0.83±0.05 | δ Sct | Constant |
| HD 169885 | 21162252 | 6.199 | 8500 | 1.55±0.09 | δ Sct | γ Dor candidate |
| HD 180347 | 298969563 | 8.204 | 7600 | 0.88±0.18 | δ Sct | Constant |
| HD 184903 | 267813367 | 7.796 | 9300 | 1.75±0.13 | δ Sct | High ampl. variation |
| HD 188103 | 209945493 | 8.103 | 9500 | 1.81±0.1 | None | High ampl. variation |
| HD 188854 | 268380299 | 7.319 | 7200 | 1.08±0.1 | δ Sct | Binary |
| HD 189574 | 172758318 | 7.523 | 7500 | 0.8±0.1 | δ Sct | Constant |
| HD 190145 | 366490533 | 7.364 | 7500 | 1.12±0.1 | δ Sct | Constant |
| HD 192662 | 380794470 | 8.647 | 8400 | 1.33±0.18 | δ Sct | High ampl. variation |
| HD 202431 | 388078603 | 7.181 | 7300 | 0.94±0.1 | δ Sct | Low ampl. variation |
| HD 210433 | 328745733 | 7.187 | 8500 | 1.89±0.05 | None | γ Dor candidate |
| HD 211643 | 331066260 | 7.039 | 8700 | 1.08±0.08 | δ Sct | δ Sct |
| HD 8251 | 248946257 | 8.170 | 8300 | 1.08±0.2 | δ Sct | Hybrid candidate |

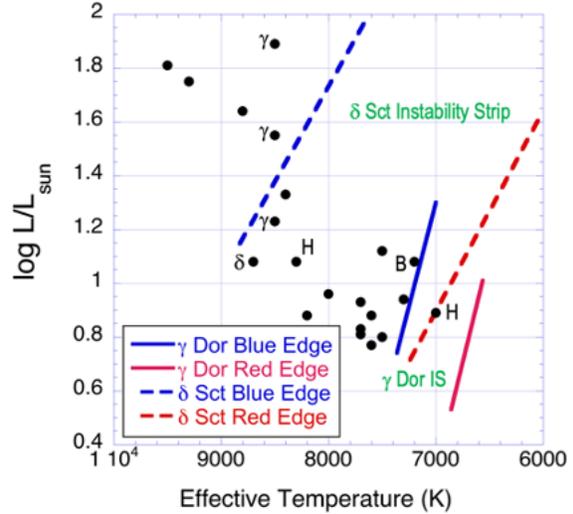

Figure 16. Location in H-R diagram of 21 Am stars observed by TESS (black dots), based on $T_{eff}$ and luminosity (L) determinations by Catanzaro et al. (2019. Also shown are the theoretical blue edge and observational red edge of the δ Sct instability strip (dashed lines) from Breger and Pamyatnykh (1998) and the theoretical red and blue edges of the γ Dor instability strip (solid lines) from Warner et al. (2003). The stars labeled are δ = δ Sct, γ = γ Dor candidate, H = hybrid candidate, and B = binary, as determined from the TESS data.

Figure 16 shows the 21 TESS objects plotted on the H-R diagram, along with the theoretical blue edge and observed red edge of the δ Sct instability strip from Breger and Pamyatnykh (1998) and the theoretical γ Dor instability strip boundaries from Warner et al. (2003). Not unexpectedly, there are many Am stars within the δ Sct instability regions that do not show pulsations. There is one δ Sct star that lies within the δ Sct IS. One of the hybrid candidates is located in the overlapping region of the two instability strips, but the other hybrid candidate is in the middle of the δ Sct IS. The three γ Dor candidates all have temperatures of 8500 K, too hot to be expected to be γ Dor stars.

Catanzaro et al. (2019) predicted, based on location in IS regions of the H-R diagram, 18 δ Sct, no γ Dor, one hybrid, and three constant or non-pulsating stars. The TESS results have in common with this prediction the one δ Sct, one hybrid, and two constant stars.

## 3. BRITE Observations

The Bright Target Explorer (BRITE) satellites were launched in 2013-2014 by a consortium of universities in Austria, Canada, and Poland. These satellites target the brightest stars of about 6[th] magnitude or less.

The five-satellite constellation includes three satellites with red filters and two with blue filters. The third blue-filter satellite, BRITE Montreal, did not separate from the upper rocket stage. The satellites are in low-Earth orbits between 600 and 800 km elevation. The BRITE fields of view are about 24° in diameter, and are observed for up to 180 days continuously for at least 15 min during each orbit. The properties of the photometric time series are constrained by the camera exposure time for bright stars, and are in the range of 1 to 5 seconds, collected about 3 to 4 times per minute, for 15 to 35 minutes per orbit.

In 2016 we proposed to observe with BRITE three B-type stars and one A-type star in the original *Kepler* field of view in the Cygnus-Lyra region. All four stars had been proposed for *Kepler* Director's Discretionary Time observations. *Kepler* data were taken for two stars, the A-type star HR 7284 in short (1-min) cadence, and the B-type star HR 7591 in long (30-min) cadence. Of the other two stars, one of them, HR 7179, has since been observed by TESS in 2-min cadence.

In addition to the puzzles surrounding the pulsations of A-type variable stars discussed in Section 2, there are also interesting questions for the more massive and luminous B-type main sequence pulsators, the β Cep and Slowly Pulsating B-type (SPB) stars and their hybrids. The β Cep stars are 7-20 times the mass of the Sun and pulsate in radial and nonradial pressure modes with periods of several hours. SPB stars are 3-9 times the mass of the Sun and pulsate in nonradial gravity modes with periods 0.5 to 5 days. See Aerts et al. (2010) and also the VSX[13] and AAVSO[14] websites for descriptions of these variable star types.

The TESS and *Kepler* data discussed below are obtained from the MAST archive. The Pre-search Data Conditioning (PDCSAP) flux with artifacts removed was used for the analysis. The amplitude spectra for the data are processed with fwpeaks[15] to determine significant frequencies and signal/noise (S/N) ratio.

### 3.1 HR 7591

HR 7591 is a B2III star with magnitude V= 5.89. It was classified as a β Cep candidate based on line profile variation observations (Telting et al. 2006). Figures 17 and 18 show the *Kepler* 30-min cadence light curve and amplitude spectrum. Some low frequency variability is evident, with the highest frequency peak at 2.25 cycles/day having a S/N ratio of 22.

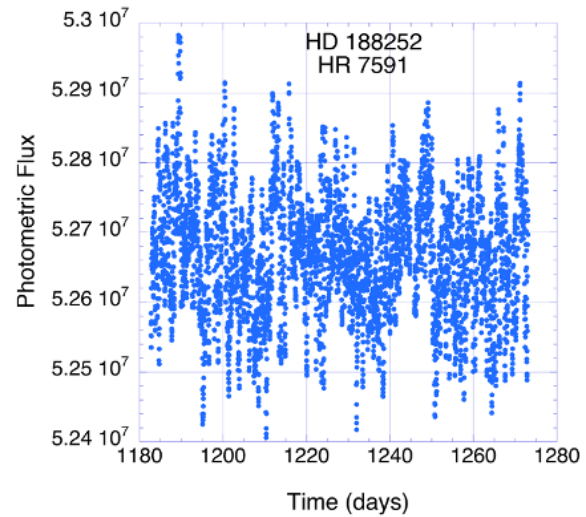

**Figure 17: HR 7591 light curve from *Kepler* 30-min cadence data.**

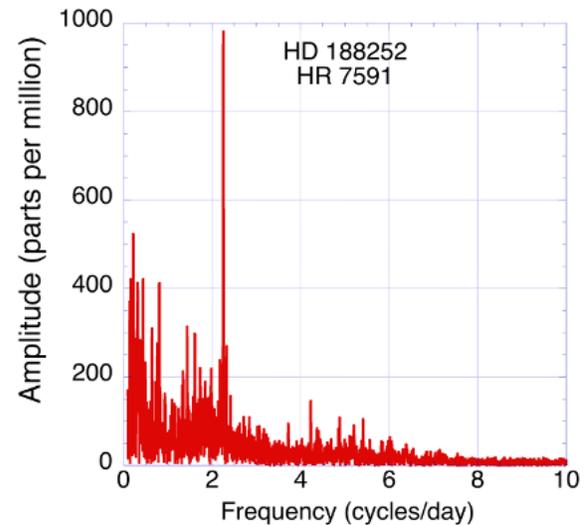

**Figure 18: HR 7591 amplitude spectrum from *Kepler* 30-min cadence data. The highest peak at 2.25 cycles/day has S/N ratio 22.**

---

[13] S. Otero, Watson, and Wils, "Variable Star Type Designations in VSX," https://www.aavso.org/vsx/index.php?view=about.vartypes

[14] "The Beta Cephei Stars and Their Relatives," https://www.aavso.org/vsots_betacep

[15] Developed and written by Waldemar Hebisch (Institute of Mathematics, Wrocław University), Zbigniew Kołaczkowski and Grzegorz Kopacki (Astronomical Institute, Wrocław University, Poland).

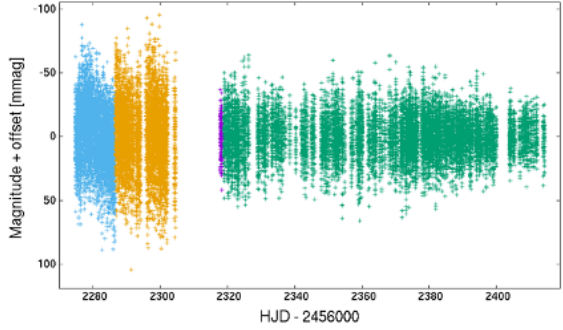

Figure 19: HR 7591 light curve data from observing runs from two BRITE satellites: BTr2 (light blue), BTr3 (orange), BHr1 (violet), and combined BHr2 and BHr3 (green).

Figure 19 shows the decorrelated BRITE light curves from the BTr (BRITE-Toronto) and BHr (BRITE-Heweliusz) setups for this star. In the figure, the colors represent data from BTr2 (light blue), BTr3 (orange), BHr1 (violet), and combined BHr2 and BHr3 (green). The BTr1 curve was very short and was removed. In the BTr observations, some instrumental effects remain because of a significant increase in CCD chip defects for this satellite, and so these data show more scatter than the BHr data. Figure 20 shows the amplitude spectrum from the BRITE data. No significant frequencies are seen, probably because the noise level near 1 millimag is comparable to the amplitude of the largest peak in the *Kepler* data (one part per thousand).

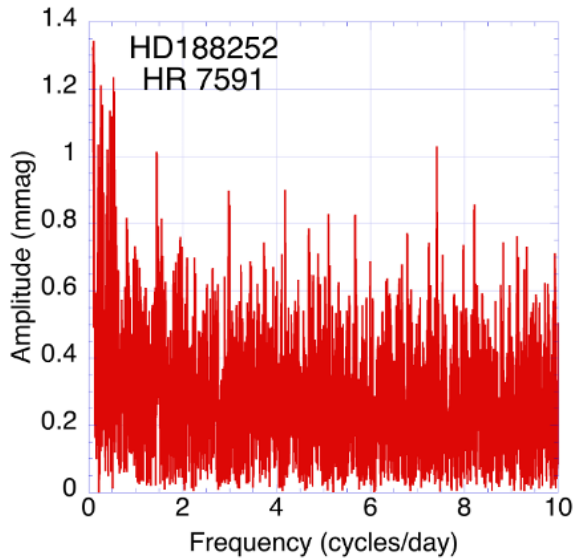

Figure 20: HR 7591 BRITE amplitude spectrum showing no significant pulsation frequencies.

## 3.2 HR 7179

HR 7179, also known as V543 Lyr, a B3V star with magnitude V=6.18, is a candidate β Cep variable (Stankov & Handler 2005). Figure 21 shows the TESS 2-min cadence light curve from sector 14; the amplitude spectrum (Fig. 22) shows low frequency SPB-like pulsations; the Fourier transform program also detects some higher frequencies (not shown) typical of β Cep pulsations with S/N level only 4-5 that may be revealed by pre-whitening the spectrum. Figure 23 shows the decorrelated and combined BRITE BHr light curves. The amplitude spectrum (Fig. 24) shows SPB-like frequencies in the same range as in the TESS data.

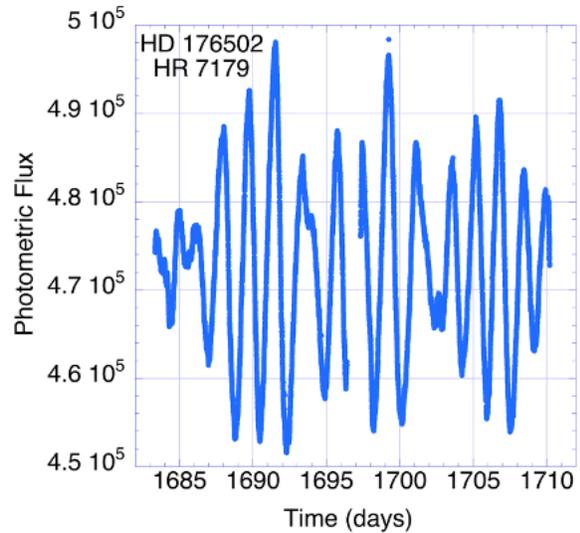

Figure 21: HR 7179 TESS 2-min cadence light curve from sector 14 observations.

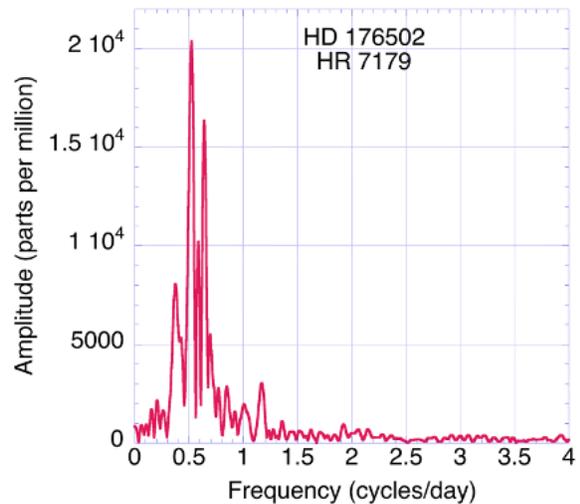

Figure 22: HR 7179 TESS amplitude spectrum showing SPB pulsations.

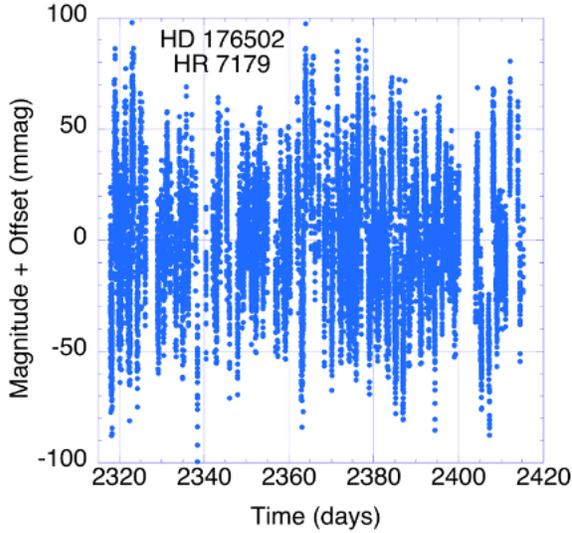

Figure 23: HR 7179 light curve from BRITE data.

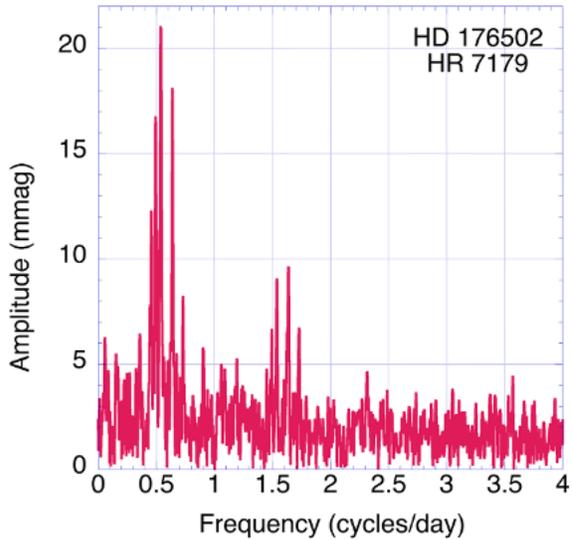

Figure 24: HR 7179 amplitude spectrum from BRITE data showing SPB pulsation frequencies similar to those seen in the TESS data.

### 3.3 HR 7403

HR 7403 is a B3Ve (emission) star with magnitude V=6.34. Be stars are named for their intermittent Hydrogen Balmer line emission thought to originate from material that has been ejected from the star; these stars may also show nonradial pulsations and rotational variability (Rivinius et al. 2013).

Figure 25 shows the decorrelated BRITE BHr light curve. Figure 26 shows the amplitude spectrum, with two significant frequencies very near 1 cycle/day (S/N 7.9) and 2 cycles/day (S/N 5.8), in the range typical for a Be star. While one might suspect that such frequencies are artifacts, multiple frequencies around 1 and 2 cycles/day also were found for this star in the *Kepler* data by Pope et al. (2019a) in their campaign to extract light curves from unsaturated halo pixels around bright stars. More work is needed to understand the origins of HR 7403's variability.

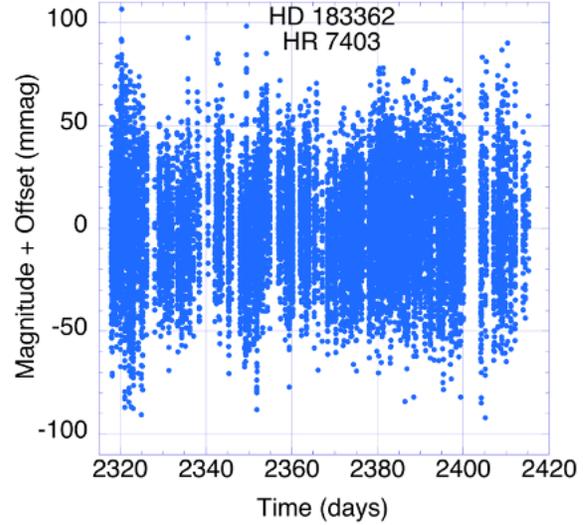

Figure 25: HR 7403 light curve from BRITE data.

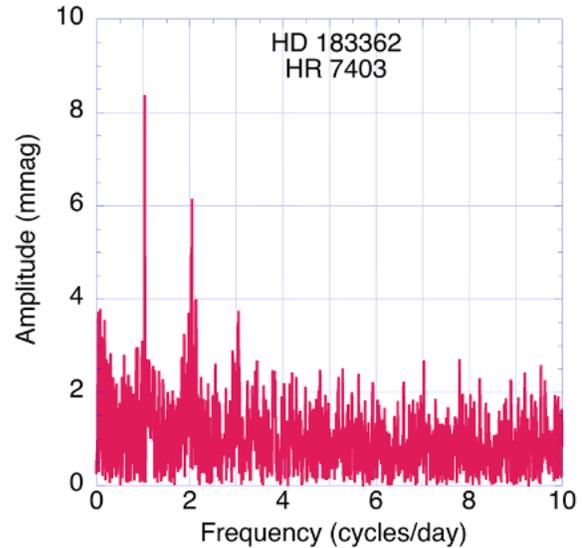

Figure 26: HR 7403 amplitude spectrum from BRITE data showing significant frequencies at around 1 and 2 cycles/day.

### 3.4 HR 7284

HR 7284 is an A3V star with magnitude V=6.18. This star has many similarities to the A-type star HD 187547, for which Antoci et al. (2014) proposed that some of the higher frequency pulsations observed in *Kepler* data are driven by turbulent pressure in the hydrogen ionization zone (see also Smalley et al. 2017). Figure 27 shows the light curve from 6 months of 1-minute cadence *Kepler* data. The Fourier analysis of this light curve (Fig. 28) shows a rich spectrum of low-amplitude pulsations in the δ Sct frequency range. Analysis is under way to determine whether HR 7284 also shows modes that could be driven by the proposed turbulent pressure mechanism.

Figures 29 and 30 show the HR 7284 BRITE BHr light curve and amplitude spectrum. The noise level in the amplitude spectrum is 0.6 to 1 mmag, too high to show the low-amplitude (<70 ppm) peaks seen in the *Kepler* data. The highest peaks in the amplitude spectrum have S/N level < 4, considered too low to be significant.

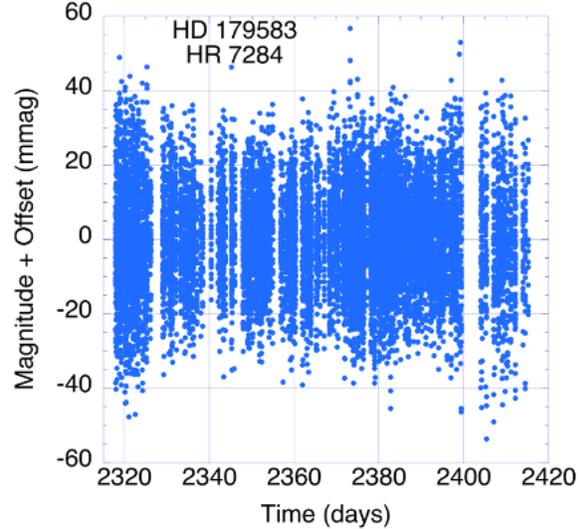
Figure 29: HR 7284 light curve from BRITE data.

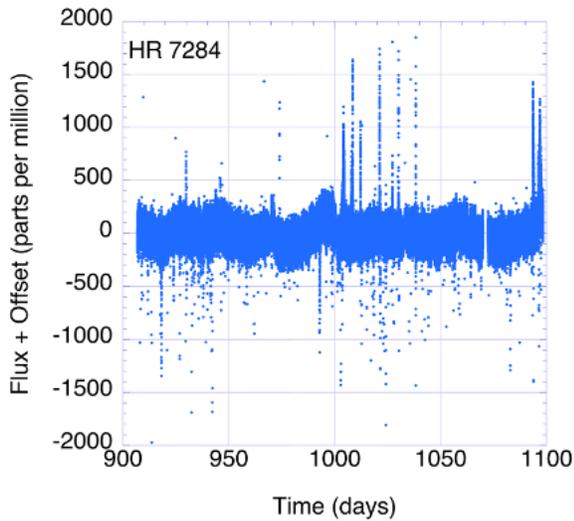
Figure 27: HR 7284 light curve from *Kepler* 1-min cadence data.

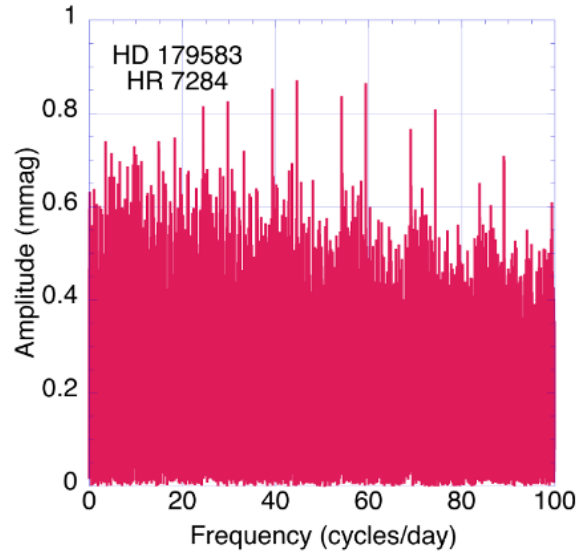
Figure 30: HR 7284 amplitude spectrum from BRITE data showing noise level at 0.6-1 mmag and no significant frequencies.

## 4. Conclusions

Of the 21 Am stars in the 62-star Catanzaro et al. sample observed by TESS so far, we find one δ Sct star and two δ Sct/γ Dor hybrid candidates. However, the stars in the sample are not necessarily expected to show δ Sct pulsations because the diffusive processes responsible for the peculiar element abundances also should deplete helium from the pulsation driving region and inhibit pulsations. There are also three γ Dor candidates that do not lie near the theoretical γ Dor instability region and six stars showing photometric variations that may or may not be caused by pulsations that require explanation.

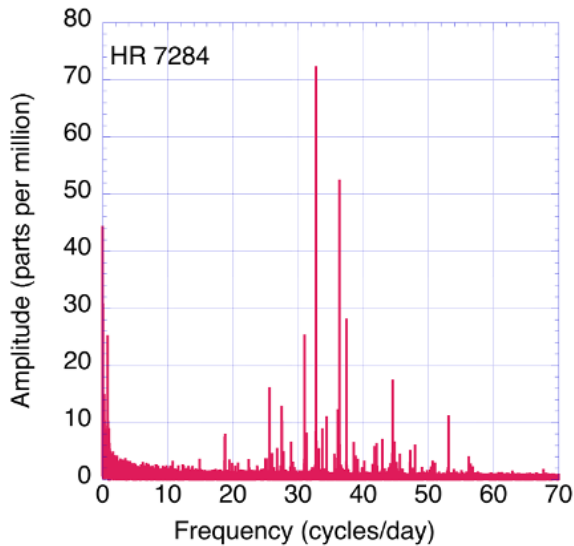
Figure 28: HR 7284 amplitude spectrum from *Kepler* data showing many frequencies in the δ Sct range.

Regarding the BRITE observations, for two stars, HR 7274 and HR 7591, prior *Kepler* and TESS observations with lower noise levels revealed low-amplitude variability that is not visible in the BRITE data. The BRITE observations did uncover low-amplitude peaks in the Be star HR 7403 that may be useful in combination with observations from other sources to further understanding of the complex processes in this type of star. BRITE observations of HR 7179 are able to detect SPB pulsations that were seen in earlier *Kepler* data.

## 5. Future Work

All of the stars discussed here are bright and amenable to ground-based photometric and spectroscopic observations, including observations by amateur observers. Some possibilities include time-series spectroscopy and multicolor photometry, useful for mode identifications and distinguishing between causes of variability, and to determine whether amplitudes or frequency content changes over time.

We could also examine the TESS 30-min cadence target pixel files from the full-frame images available for many of the stars. In addition, the *Kepler* full-frame images, including haloes around saturated pixels, could be mined for variability information. Pope et al. (2019a,b) have already used such data to reconstruct light curves for hundreds of stars, including HR 7403 as discussed above, that were too bright to have been targeted by *Kepler*.

These data will provide key constraints for stellar evolution and pulsation modeling studies, and for revealing stellar interior structure and processes via asteroseismology.

## 6. Acknowledgements

We are grateful for data from the TESS Guest Investigator Cycle 1 and 2 programs, the BRITE Constellation satellites, and the *Kepler* Director's Discretionary Time program.

JG gratefully acknowledges a Los Alamos National Laboratory Center for Space and Earth Sciences grant CSES XWPB ARR0GZIK. APi acknowledges support from the NCN grant no. 2016/21/B/ST9/01126. PG acknowledges funding from the German Aerospace Center (Deutsches Zentrum für Luft- und Raumfahrt) under PLATO Data Center grant 50OO1501. GH acknowledges financial support by the Polish National Science Centre (NCN) under grant 2015/18/A/ST9/00578.

This research has made use of the SIMBAD database, operated at CDS, Strasbourg, France, the Mikulski Archive for Space Telescopes (MAST), and the TESS field of view search tool of the TESS Asteroseismic Science Consortium.